\documentclass[acmlarge,screen]{acmart}

\AtBeginDocument{%
  \providecommand\BibTeX{{%
    \normalfont B\kern-0.5em{\scshape i\kern-0.25em b}\kern-0.8em\TeX}}}

\setcopyright{none}
\copyrightyear{}
\acmYear{}
\acmDOI{}

\acmConference[CHI 2024 Workshop on HCI and Aging]{Make sure to enter the correct
  conference title from your rights confirmation emai}{May,
  2024}{Honululu, HI}
\acmBooktitle{HCI and Aging: New Directions, New Principles Workshop at ACM SIGCHI 2024} 
\acmISBN{}

\definecolor{DarkBlue}{HTML}{00008B}
\definecolor{brickred}{HTML}{f03b20}
\definecolor{dpink}{HTML}{CD1076}

\newif{\ifhidecomments}
  \hidecommentsfalse 
\ifhidecomments
    \newcommand{\sidharth}[1]{}
    \newcommand{\abhay}[1]{}
    \newcommand{\koustuv}[1]{}
    \newcommand{\ravi}[1]{}
\else
    \newcommand{\sidharth}[1]{\textbf{\small\sffamily{\textcolor{DarkBlue}{[#1 -- Sidharth]}}}}
    \newcommand{\abhay}[1]{\textbf{\small\sffamily{\textcolor{orange}{[#1 -- Abhay]}}}}
    \newcommand{\ravi}[1]{\textbf{\small\sffamily{\textcolor{brickred}{[#1 -- Ravi]}}}}
    \newcommand{\koustuv}[1]{\textbf{\small\sffamily{\textcolor{dpink}{[#1 -- Koustuv]}}}}
  \fi

\begin{document}

\title{Exploring the Role of LLMs for Supporting Older Adults:\\ Opportunities and Concerns}

\author{Sidharth Kaliappan}
\authornote{Both authors contributed equally to this research.}
\email{skaliappan@umass.edu}
\author{Abhay Sheel Anand}
\authornotemark[1]
\email{asanand@umass.edu}
\affiliation{%
  \institution{University of Massachusetts Amherst}
  \country{USA}
}

\author{Koustuv Saha}
\affiliation{%
  \institution{University of Illinois Urbana-Champaign}
  \country{USA}
}

\author{Ravi Karkar}
\affiliation{%
  \institution{University of Massachusetts Amherst}
  \country{USA}
}







\begin{abstract}

In this position paper, we explore the potential role of large language models (LLMs) in enhancing healthcare and quality of life for older adults. 
We explore some of the existing research in HCI around technology for older adults and examine the role of LLMs in enhancing it.
We also discuss the digital divide and emphasize the need for inclusive technology design. 
At the same time, we also surface concerns regarding privacy, security, and the accuracy of information provided by LLMs, alongside the importance of user-centered design to make technology accessible and effective for the elderly. 
We show the transformative possibilities of LLM-supported interactions at the intersection of aging, technology, and human-computer interaction, advocating for further research and development in this area.

\end{abstract}





\maketitle

\section{Introduction}
With the aging global population, there is an urgent need for technology support---both medical and in terms of quality of life---for older adults~\cite{mahishale2015ageing}. 
Within HCI, researchers have studied the design of technology to better meet the evolving needs to older adults~\cite{pradhan2020understanding, pang2021technology}.
The COVID-19 pandemic has surfaced the need for technology as a basic necessity not just for entertainment and staying connected but, for more critical needs such as healthcare~\cite{xie2021going}.


HCI researchers have made significant technological advances to address diverse needs of older adults and their caregivers. 
From the development of intuitive user interfaces that accommodate declining sensory and motor functions to the creation of social robots that mitigate feelings of loneliness and isolation~\cite{kubota2022cognitively, mihajlov2015intuitive}. 
A few instances of how HCI has worked to enhance the quality of life for the elderly include wearable technology that tracks health metrics in real-time, easily accessible digital platforms that promote social interaction, and smart home technologies that let senior citizens live independently for longer~\cite{chen2023digital, ollevier2020can, lee2022technology}.

In light of the groundbreaking advances in artificial intelligence provided by the recent large language models~\cite{gpt4, bard}, we deem it timely to revisit the past HCI research on technology design for older adults and view it through a new lens of LLMs.
Because of their large capacity for producing, comprehending, and interacting with human language, LLMs have the potential to transform how we interact with technology~\cite{hamalainen2023evaluating}.
Modern LLMs like OpenAI's GPT-4~\cite{gpt4} and Google's Bard Gemini~\cite{bard} exhibit enhanced performance over previous models due to significant improvements in architecture, training datasets, and learning algorithms. 
These advancements result in a better understanding of context, a more accurate generation of text, and an improved ability to handle complex queries, making them more effective and versatile in a wide range of applications~\cite{naveed2023comprehensive}.
Although there are issues with LLMs that we need to address too (e.g inaccurate or irrelevant information~\cite{li2023halueval, rambhatlamaking}), it is worthwhile to explore what value they might contribute to this domain.
The goal of this article is to examine the various ways that LLM-enabled interactions can improve healthcare and quality of life for older adults. 
By revisiting previous studies and projects in HCI with a focus on the integration of LLM technologies, we aim to highlight the transformative possibilities that lie at the intersection of aging, technology, and human-computer interaction.

The premise behind this article is that the combination of HCI research and LLM technologies has the potential to produce new advancements in assistive technologies.
The potential uses of LLMs to improve the lives of older adults are numerous, ranging from individualized virtual assistants that provide company and assistance with everyday tasks to advanced platforms that offer specialized health advice and cognitive stimulation.

\section{Background}


\subsection{Access to Technology for Older Adults}


The body of research on technology's impact on older adults aims to enhance their well-being, independence, and social connections, particularly through their increasing engagement with digital and wearable devices. In recent times, the integration of technology in healthcare has been significantly amplified, not only by the COVID-19 pandemic~\cite{sixsmith2022older} but also by the growing need for remote technologies and advances in generative AI technologies.
The initial excitement about technology's potential is tempered by the noticeable digital divide affecting older adults, emphasizing the need for inclusive access.
This divide becomes apparent when considering that the effectiveness of tailored instructional methods in enhancing technology adoption among older adults indirectly highlights the existing barriers to access and use~\cite{kim2022exploring}. In particular, wearable devices stand out as a promising tool in eldercare. Still, their acceptance is contingent upon factors such as cost, performance expectations, and an individual's openness to innovation~\cite{chen2023research}. Challenges persist, including the need to address ageist attitudes~\cite{mannheim2023ageism} and the absence of inclusive design in digital health~\cite{pang2021technology}. Moreover, there's an evident lack of representation of older adults, especially those with cognitive impairments, in technology research~\cite{mace2022older, guu2023wearable}.

\subsection{Technology for Smart Health Monitoring}
Bridging the digital divide also means integrating smart health technology into the daily lives of older adults. Smart health technology focuses on using the integration of wearable devices, AI, and smart home tech to support older adults and enable aging in place~\cite{facchinetti2023can, kim2022home}. These innovations aim to enhance chronic condition management and healthcare by offering real-time monitoring, early detection, and intervention~\cite{sabry2022machine}. Smart home and in-home monitoring technologies are essential for independence and health tracking, while the integration of machine learning into wearables extends diagnostic capabilities but raises privacy and user acceptance challenges~\cite{cristiano2022older}. The development of smart health platforms requires a user-centered design and interdisciplinary collaboration for effective implementation~\cite{palanisamy2023remote, wang2023integrating}. IoT sensors and AI enable personalized care through real-time data in long-term settings but face challenges in health customization, access to information, remote care, and telehealth integration, highlighting the need for more research in healthcare technology design for older adults.

\subsection{Digital Literacy and Security}
In a hyper-connected society, digital literacy has significant impact on a person's quality of life. 
From access to information, staying connected, to actively participating in society, digital literacy is a vital skill for older adults.
As smart health technologies advance, equally critical is empowering older adults with the digital literacy to effectively use these tools. 
Additionally, digital literacy is essential for navigating online risks such as misinformation and security threats. 
Tailored educational initiatives are crucial to counter emotionally-driven misinformation, emphasizing the need for nuanced protective strategies~\cite{vitak2020trust,pasquetto2020tackling}.

Amidst the COVID-19 pandemic, the importance of enhancing health and data literacy for older adults has become more apparent, highlighting the urgent need for interventions that bolster their digital competencies~\cite{moore2022digital}. 
Despite progress, obstacles such as susceptibility to scams and difficulties in identifying misinformation persist, indicating a pressing need for comprehensive measures to improve digital literacy and resilience among older adults~\cite{bmj}.

\section{Discussion}


Technology is advancing, particularly in healthcare systems, and it's becoming more accessible to older adults~\cite{chen2023digital}. 
However, there are clear barriers to this progress.
These barriers are predominantly characterized by a lack of understanding of technical features, high cost of technology, confusing instructions, and privacy concerns~\cite{harris2022older,wilson2021barriers}. 
Modern LLMs change how the technology interfaces with  users, offering the potential to overcome these barriers. 
We highlight some areas relevant to aging where we see LLMs have a big impact over the coming years.
In this section we refer to the broader technology as LLMs even though its instantiation in the examples below can be in the form of chatbots, smart assistants, or other modalities.

\subsubsection*{LLMs for Accessibility}
Modern LLMs have the capability to facilitate natural conversation, this represents an advancement over traditional search engines. 
Unlike the latter, which often require specific query phrasing and give results that require further navigation and interpretation, LLMs allow for a more intuitive and natural user experience. 
This ability lowers the threshold of required technological knowledge to effectively engage with and leverage LLM-based querying.
Having the ability to maintain continuity in conversation enhances the accessibility of information and support.
From an HCI perspective, this shift towards conversational interfaces provides a simpler mental model for users~\cite{wang2021towards}. 
This moves the focus from needing to understand and manipulate technology to accomplish tasks, to directly fulfill those tasks by conversing.
This change would increase use and utility by reducing the barriers traditionally associated with technology use among older adults.
Although reliance on LLMs could raise concerns about user privacy and data security, as these systems often need to process sensitive personal information to deliver customized responses.

\subsubsection*{LLMs for Health Monitoring}

LLMs can play a supportive role in healthcare, especially for older adults, by augmenting the capabilities of clinicians and healthcare providers.
Programming LLMs to comprehend and respond to health-related inquiries can offer immediate, conversational access to medical information, including medication reminders and initial diagnostic guidance.
They can also play a crucial role in monitoring dietary habits, suggesting adjustments to improve nutrition in accordance with specific health goals or conditions such as diabetes or hypertension~\cite{limketkai2021age}.
Under the supervision of clinicians, LLMs can offer valuable support in promoting physical health by customizing and suggesting exercise routines that align with the specific abilities and health goals of individuals, fostering a more active and healthier lifestyle.
These models can also serve as crucial intermediaries for professional healthcare services, skillfully identifying symptoms that may require medical intervention and guiding individuals towards seeking consultations with healthcare providers~\cite{gudala2022benefits}.
But integration of LLMs in healthcare can introduces potential risks of misdiagnosis or misinformation, especially if the models are not accurately updated or tailored to individual patient contexts, which could lead to adverse health outcomes

\subsubsection*{LLMs for Digital Literacy and Security}
By facilitating natural, conversational dialogues, LLMs can play a crucial role in assisting older adults to familiarize themselves with digital tools and platforms.
These models can guide users from grasping basic functions, such as sending emails, to navigating more intricate tasks like online banking and digital shopping. 
Enhancing the design of LLMs such as improvements in user interfaces, voice-activated commands for ease of use, and the provision of immediate, constructive feedback can significantly elevate the learning journey. 
Incorporating personalization and adaptive learning helps cater to varying levels of digital literacy, while embedded support systems and visual aids enhance usability.
\\
As this demographic is often targeted by fraudulent schemes, LLMs can protect older adults by educating and warning users about common scams and malicious tactics. 
For instance, an LLM could be programmed to recognize and flag communications that bear the hallmarks of phishing emails, such as requests for sensitive information under the guise of bank or government officials.
Through conversational interactions, these models can explain the signs of phishing emails, and suspicious messages. 
Providing older adults with the knowledge to recognize and avoid potential threats.
Additionally, LLMs can offer 24/7 technical support, guiding users through securing their personal information and navigating privacy settings across various platforms. 
This constant availability ensures that older adults can receive immediate assistance whenever they encounter security concerns, making technology use safer and more reassuring
While LLMs bring benefits in user engagement and services but pose privacy and security risks, as sensitive information shared with them could lead to data breaches or misuse. 
Protecting this data requires strong encryption, strict access controls, and clear data policies to avoid unauthorized use.

\subsubsection*{Known Concerns with LLMs}
Despite these advancements, LLMs are not without drawbacks. 
One of the primary concerns is the accuracy of the information provided. 
LLMs can still produce inaccurate information or "hallucinate" responses, leading to potential irrelevant or misinformation~\cite{li2023halueval, rambhatlamaking}. 
The problem is compounded by LLM outputs not showing any measure of confidence.
This means that both \textit{true} and \textit{false} information is presented with the same level of confidence.
The length of the contex window\cite{liu2024lost} in language models is also active concern in the community where the relevant information is often lost in the middle of long contexts.
Furthermore, the effective use of LLMs still requires a degree of "prompt engineering," or the ability to phrase queries in a manner that yields the most accurate and relevant responses. 
While LLMs have made significant progress in understanding natural language, the precision of user prompts can greatly influence the quality of the interaction.

Additionally, there are concerns regarding privacy and data security. As LLMs often require access to personal information to provide tailored advice, there's a risk of sensitive data being inadvertently exposed or misused. Ensuring the confidentiality and security of user data is essential. It still remains a significant challenge in the deployment of LLMs for health-related applications.

\section{Conclusion}
Our goal with the position piece is not to provide a complete vision but, to instead point to an emerging direction of how technology is evolving and provide some examples of how we imagine it may be relevant to this workshop. 
LLMs in their current incarnation are still a nascent technology with huge potential and equally big concerns of potential harm. 
As HCI researchers, we are well positioned to critically examine this technology, particularly as to its application to older adults. 
What kind of scaffolding should be provided when introducing LLMs to tech for older adults? 
What new interactions can now be accomplished with minimal user input that can enhance the quality of life of older adults? 
How do we explain the new threat models (e.g., hallucinations) to people who might not be \textit{tech savvy}?

We hope to participate in the workshop to discuss these questions and to engage with other researchers around what they are exploring in the space. 

    


\bibliographystyle{ACM-Reference-Format}
\bibliography{sample-authordraft}


\begin{thebibliography}{37}


\ifx \showCODEN    \undefined \def \showCODEN     #1{\unskip}     \fi
\ifx \showDOI      \undefined \def \showDOI       #1{#1}\fi
\ifx \showISBNx    \undefined \def \showISBNx     #1{\unskip}     \fi
\ifx \showISBNxiii \undefined \def \showISBNxiii  #1{\unskip}     \fi
\ifx \showISSN     \undefined \def \showISSN      #1{\unskip}     \fi
\ifx \showLCCN     \undefined \def \showLCCN      #1{\unskip}     \fi
\ifx \shownote     \undefined \def \shownote      #1{#1}          \fi
\ifx \showarticletitle \undefined \def \showarticletitle #1{#1}   \fi
\ifx \showURL      \undefined \def \showURL       {\relax}        \fi
\providecommand\bibfield[2]{#2}
\providecommand\bibinfo[2]{#2}
\providecommand\natexlab[1]{#1}
\providecommand\showeprint[2][]{arXiv:#2}

\bibitem[bar(2024)]%
        {bard}
 \bibinfo{year}{2024}\natexlab{}.
\newblock \bibinfo{booktitle}{\emph{Google Bard Gemini}}.
\newblock
\urldef\tempurl%
\url{https://gemini.google.com/}
\showURL{%
Retrieved February 20, 2023 from \tempurl}


\bibitem[gpt(2024)]%
        {gpt4}
 \bibinfo{year}{2024}\natexlab{}.
\newblock \bibinfo{booktitle}{\emph{OpenAI GPT-4}}.
\newblock
\urldef\tempurl%
\url{https://openai.com/research/gpt-4}
\showURL{%
Retrieved February 20, 2023 from \tempurl}


\bibitem[Chen et~al\mbox{.}(2023a)]%
        {chen2023digital}
\bibfield{author}{\bibinfo{person}{Chuanrui Chen}, \bibinfo{person}{Shichao Ding}, {and} \bibinfo{person}{Joseph Wang}.} \bibinfo{year}{2023}\natexlab{a}.
\newblock \showarticletitle{Digital health for aging populations}.
\newblock \bibinfo{journal}{\emph{Nature Medicine}} \bibinfo{volume}{29}, \bibinfo{number}{7} (\bibinfo{year}{2023}), \bibinfo{pages}{1623--1630}.
\newblock


\bibitem[Chen et~al\mbox{.}(2023b)]%
        {chen2023research}
\bibfield{author}{\bibinfo{person}{Junxun Chen}, \bibinfo{person}{Tao Wang}, \bibinfo{person}{Zhenyu Fang}, {and} \bibinfo{person}{Hongtao Wang}.} \bibinfo{year}{2023}\natexlab{b}.
\newblock \showarticletitle{Research on elderly users' intentions to accept wearable devices based on the improved UTAUT model}.
\newblock \bibinfo{journal}{\emph{Frontiers in Public Health}}  \bibinfo{volume}{10} (\bibinfo{year}{2023}), \bibinfo{pages}{1035398}.
\newblock


\bibitem[Cristiano et~al\mbox{.}(2022)]%
        {cristiano2022older}
\bibfield{author}{\bibinfo{person}{Alessia Cristiano}, \bibinfo{person}{Stela Musteata}, \bibinfo{person}{Sara De~Silvestri}, \bibinfo{person}{Valerio Bellandi}, \bibinfo{person}{Paolo Ceravolo}, \bibinfo{person}{Matteo Cesari}, \bibinfo{person}{Domenico Azzolino}, \bibinfo{person}{Alberto Sanna}, {and} \bibinfo{person}{Diana Trojaniello}.} \bibinfo{year}{2022}\natexlab{}.
\newblock \showarticletitle{Older adults’ and clinicians’ perspectives on a smart health platform for the aging population: Design and evaluation study}.
\newblock \bibinfo{journal}{\emph{JMIR aging}} \bibinfo{volume}{5}, \bibinfo{number}{1} (\bibinfo{year}{2022}), \bibinfo{pages}{e29623}.
\newblock


\bibitem[Facchinetti et~al\mbox{.}(2023)]%
        {facchinetti2023can}
\bibfield{author}{\bibinfo{person}{Gabriella Facchinetti}, \bibinfo{person}{Giorgia Petrucci}, \bibinfo{person}{Beatrice Albanesi}, \bibinfo{person}{Maria~Grazia De~Marinis}, {and} \bibinfo{person}{Michela Piredda}.} \bibinfo{year}{2023}\natexlab{}.
\newblock \showarticletitle{Can smart home technologies help older adults manage their chronic condition? A systematic literature review}.
\newblock \bibinfo{journal}{\emph{International Journal of Environmental Research and Public Health}} \bibinfo{volume}{20}, \bibinfo{number}{2} (\bibinfo{year}{2023}), \bibinfo{pages}{1205}.
\newblock


\bibitem[Gudala et~al\mbox{.}(2022)]%
        {gudala2022benefits}
\bibfield{author}{\bibinfo{person}{Meghana Gudala}, \bibinfo{person}{Mary Ellen~Trail Ross}, \bibinfo{person}{Sunitha Mogalla}, \bibinfo{person}{Mandi Lyons}, \bibinfo{person}{Padmavathy Ramaswamy}, \bibinfo{person}{Kirk Roberts}, {et~al\mbox{.}}} \bibinfo{year}{2022}\natexlab{}.
\newblock \showarticletitle{Benefits of, barriers to, and needs for an artificial intelligence--powered medication information voice chatbot for older adults: interview study with geriatrics experts}.
\newblock \bibinfo{journal}{\emph{JMIR aging}} \bibinfo{volume}{5}, \bibinfo{number}{2} (\bibinfo{year}{2022}), \bibinfo{pages}{e32169}.
\newblock


\bibitem[Guu et~al\mbox{.}(2023)]%
        {guu2023wearable}
\bibfield{author}{\bibinfo{person}{Ta-Wei Guu}, \bibinfo{person}{Marijn Muurling}, \bibinfo{person}{Zunera Khan}, \bibinfo{person}{Chris Kalafatis}, \bibinfo{person}{Dag Aarsland}, \bibinfo{person}{Anna-Katharine Brem}, {et~al\mbox{.}}} \bibinfo{year}{2023}\natexlab{}.
\newblock \showarticletitle{Wearable devices: underrepresentation in the ageing society}.
\newblock \bibinfo{journal}{\emph{The Lancet Digital Health}} \bibinfo{volume}{5}, \bibinfo{number}{6} (\bibinfo{year}{2023}), \bibinfo{pages}{e336--e337}.
\newblock


\bibitem[H{\"a}m{\"a}l{\"a}inen et~al\mbox{.}(2023)]%
        {hamalainen2023evaluating}
\bibfield{author}{\bibinfo{person}{Perttu H{\"a}m{\"a}l{\"a}inen}, \bibinfo{person}{Mikke Tavast}, {and} \bibinfo{person}{Anton Kunnari}.} \bibinfo{year}{2023}\natexlab{}.
\newblock \showarticletitle{Evaluating large language models in generating synthetic hci research data: a case study}. In \bibinfo{booktitle}{\emph{Proceedings of the 2023 CHI Conference on Human Factors in Computing Systems}}. \bibinfo{pages}{1--19}.
\newblock


\bibitem[Harris et~al\mbox{.}(2022)]%
        {harris2022older}
\bibfield{author}{\bibinfo{person}{Maurita~T Harris}, \bibinfo{person}{Kenneth~A Blocker}, {and} \bibinfo{person}{Wendy~A Rogers}.} \bibinfo{year}{2022}\natexlab{}.
\newblock \showarticletitle{Older adults and smart technology: facilitators and barriers to use}.
\newblock \bibinfo{journal}{\emph{Frontiers in Computer Science}}  \bibinfo{volume}{4} (\bibinfo{year}{2022}), \bibinfo{pages}{835927}.
\newblock


\bibitem[Kim et~al\mbox{.}(2022a)]%
        {kim2022home}
\bibfield{author}{\bibinfo{person}{Daejin Kim}, \bibinfo{person}{Hongyi Bian}, \bibinfo{person}{Carl~K Chang}, \bibinfo{person}{Liang Dong}, \bibinfo{person}{Jennifer Margrett}, {et~al\mbox{.}}} \bibinfo{year}{2022}\natexlab{a}.
\newblock \showarticletitle{In-home monitoring technology for aging in place: scoping review}.
\newblock \bibinfo{journal}{\emph{Interactive journal of medical research}} \bibinfo{volume}{11}, \bibinfo{number}{2} (\bibinfo{year}{2022}), \bibinfo{pages}{e39005}.
\newblock


\bibitem[Kim et~al\mbox{.}(2022b)]%
        {kim2022exploring}
\bibfield{author}{\bibinfo{person}{Sunyoung Kim}, \bibinfo{person}{Willow Yao}, {and} \bibinfo{person}{Xiaotong Du}.} \bibinfo{year}{2022}\natexlab{b}.
\newblock \showarticletitle{Exploring older adults’ adoption and use of a tablet computer during COVID-19: longitudinal qualitative study}.
\newblock \bibinfo{journal}{\emph{JMIR aging}} \bibinfo{volume}{5}, \bibinfo{number}{1} (\bibinfo{year}{2022}), \bibinfo{pages}{e32957}.
\newblock


\bibitem[Kubota et~al\mbox{.}(2022)]%
        {kubota2022cognitively}
\bibfield{author}{\bibinfo{person}{Alyssa Kubota}, \bibinfo{person}{Dagoberto Cruz-Sandoval}, \bibinfo{person}{Soyon Kim}, \bibinfo{person}{Elizabeth~W Twamley}, {and} \bibinfo{person}{Laurel~D Riek}.} \bibinfo{year}{2022}\natexlab{}.
\newblock \showarticletitle{Cognitively assistive robots at home: Hri design patterns for translational science}. In \bibinfo{booktitle}{\emph{2022 17th ACM/IEEE International Conference on Human-Robot Interaction (HRI)}}. IEEE, \bibinfo{pages}{53--62}.
\newblock


\bibitem[Lee(2022)]%
        {lee2022technology}
\bibfield{author}{\bibinfo{person}{Chaiwoo Lee}.} \bibinfo{year}{2022}\natexlab{}.
\newblock \showarticletitle{Technology and aging: the jigsaw puzzle of design, development and distribution}.
\newblock \bibinfo{journal}{\emph{Nature Aging}} \bibinfo{volume}{2}, \bibinfo{number}{12} (\bibinfo{year}{2022}), \bibinfo{pages}{1077--1079}.
\newblock


\bibitem[Li et~al\mbox{.}(2023)]%
        {li2023halueval}
\bibfield{author}{\bibinfo{person}{Junyi Li}, \bibinfo{person}{Xiaoxue Cheng}, \bibinfo{person}{Xin Zhao}, \bibinfo{person}{Jian-Yun Nie}, {and} \bibinfo{person}{Ji-Rong Wen}.} \bibinfo{year}{2023}\natexlab{}.
\newblock \showarticletitle{Halueval: A large-scale hallucination evaluation benchmark for large language models}. In \bibinfo{booktitle}{\emph{The 2023 Conference on Empirical Methods in Natural Language Processing}}.
\newblock


\bibitem[Limketkai et~al\mbox{.}(2021)]%
        {limketkai2021age}
\bibfield{author}{\bibinfo{person}{Berkeley~N Limketkai}, \bibinfo{person}{Kasuen Mauldin}, \bibinfo{person}{Natalie Manitius}, \bibinfo{person}{Laleh Jalilian}, {and} \bibinfo{person}{Bradley~R Salonen}.} \bibinfo{year}{2021}\natexlab{}.
\newblock \showarticletitle{The age of artificial intelligence: use of digital technology in clinical nutrition}.
\newblock \bibinfo{journal}{\emph{Current surgery reports}} \bibinfo{volume}{9}, \bibinfo{number}{7} (\bibinfo{year}{2021}), \bibinfo{pages}{20}.
\newblock


\bibitem[Liu et~al\mbox{.}(2024)]%
        {liu2024lost}
\bibfield{author}{\bibinfo{person}{Nelson~F Liu}, \bibinfo{person}{Kevin Lin}, \bibinfo{person}{John Hewitt}, \bibinfo{person}{Ashwin Paranjape}, \bibinfo{person}{Michele Bevilacqua}, \bibinfo{person}{Fabio Petroni}, {and} \bibinfo{person}{Percy Liang}.} \bibinfo{year}{2024}\natexlab{}.
\newblock \showarticletitle{Lost in the middle: How language models use long contexts}.
\newblock \bibinfo{journal}{\emph{Transactions of the Association for Computational Linguistics}}  \bibinfo{volume}{12} (\bibinfo{year}{2024}), \bibinfo{pages}{157--173}.
\newblock


\bibitem[Mace et~al\mbox{.}(2022)]%
        {mace2022older}
\bibfield{author}{\bibinfo{person}{Ryan~A Mace}, \bibinfo{person}{Meghan~K Mattos}, {and} \bibinfo{person}{Ana-Maria Vranceanu}.} \bibinfo{year}{2022}\natexlab{}.
\newblock \showarticletitle{Older adults can use technology: why healthcare professionals must overcome ageism in digital health}.
\newblock \bibinfo{journal}{\emph{Translational Behavioral Medicine}} \bibinfo{volume}{12}, \bibinfo{number}{12} (\bibinfo{year}{2022}), \bibinfo{pages}{1102--1105}.
\newblock


\bibitem[Mahishale(2015)]%
        {mahishale2015ageing}
\bibfield{author}{\bibinfo{person}{Vinay Mahishale}.} \bibinfo{year}{2015}\natexlab{}.
\newblock \showarticletitle{Ageing world: Health care challenges}.
\newblock \bibinfo{journal}{\emph{Journal of the Scientific Society}} \bibinfo{volume}{42}, \bibinfo{number}{3} (\bibinfo{year}{2015}), \bibinfo{pages}{138--143}.
\newblock


\bibitem[Mannheim et~al\mbox{.}(2023)]%
        {mannheim2023ageism}
\bibfield{author}{\bibinfo{person}{Ittay Mannheim}, \bibinfo{person}{Eveline~JM Wouters}, \bibinfo{person}{Hanna K{\"o}ttl}, \bibinfo{person}{Leonieke~C van Boekel}, \bibinfo{person}{Rens Brankaert}, {and} \bibinfo{person}{Yvonne van Zaalen}.} \bibinfo{year}{2023}\natexlab{}.
\newblock \showarticletitle{Ageism in the discourse and practice of designing digital technology for older persons: A scoping review}.
\newblock \bibinfo{journal}{\emph{The Gerontologist}} \bibinfo{volume}{63}, \bibinfo{number}{7} (\bibinfo{year}{2023}), \bibinfo{pages}{1188--1200}.
\newblock


\bibitem[Mihajlov et~al\mbox{.}(2015)]%
        {mihajlov2015intuitive}
\bibfield{author}{\bibinfo{person}{Martin Mihajlov}, \bibinfo{person}{Effie Lai-Chong Law}, {and} \bibinfo{person}{Mark Springett}.} \bibinfo{year}{2015}\natexlab{}.
\newblock \showarticletitle{Intuitive learnability of touch gestures for technology-na{\"\i}ve older adults}.
\newblock \bibinfo{journal}{\emph{Interacting with Computers}} \bibinfo{volume}{27}, \bibinfo{number}{3} (\bibinfo{year}{2015}), \bibinfo{pages}{344--356}.
\newblock


\bibitem[Moore and Hancock(2022)]%
        {moore2022digital}
\bibfield{author}{\bibinfo{person}{Ryan~C Moore} {and} \bibinfo{person}{Jeffrey~T Hancock}.} \bibinfo{year}{2022}\natexlab{}.
\newblock \showarticletitle{A digital media literacy intervention for older adults improves resilience to fake news}.
\newblock \bibinfo{journal}{\emph{Scientific reports}} \bibinfo{volume}{12}, \bibinfo{number}{1} (\bibinfo{year}{2022}), \bibinfo{pages}{6008}.
\newblock


\bibitem[Naveed et~al\mbox{.}(2023)]%
        {naveed2023comprehensive}
\bibfield{author}{\bibinfo{person}{Humza Naveed}, \bibinfo{person}{Asad~Ullah Khan}, \bibinfo{person}{Shi Qiu}, \bibinfo{person}{Muhammad Saqib}, \bibinfo{person}{Saeed Anwar}, \bibinfo{person}{Muhammad Usman}, \bibinfo{person}{Nick Barnes}, {and} \bibinfo{person}{Ajmal Mian}.} \bibinfo{year}{2023}\natexlab{}.
\newblock \showarticletitle{A comprehensive overview of large language models}.
\newblock \bibinfo{journal}{\emph{arXiv preprint arXiv:2307.06435}} (\bibinfo{year}{2023}).
\newblock


\bibitem[Ollevier et~al\mbox{.}(2020)]%
        {ollevier2020can}
\bibfield{author}{\bibinfo{person}{Aline Ollevier}, \bibinfo{person}{Gabriel Aguiar}, \bibinfo{person}{Marco Palomino}, {and} \bibinfo{person}{Ingeborg~Sylvia Simpelaere}.} \bibinfo{year}{2020}\natexlab{}.
\newblock \showarticletitle{How can technology support ageing in place in healthy older adults? A systematic review}.
\newblock \bibinfo{journal}{\emph{Public health reviews}}  \bibinfo{volume}{41} (\bibinfo{year}{2020}), \bibinfo{pages}{1--12}.
\newblock


\bibitem[Palanisamy et~al\mbox{.}(2023)]%
        {palanisamy2023remote}
\bibfield{author}{\bibinfo{person}{Preethi Palanisamy}, \bibinfo{person}{Amudhavalli Padmanabhan}, \bibinfo{person}{Asokan Ramasamy}, {and} \bibinfo{person}{Sakthivel Subramaniam}.} \bibinfo{year}{2023}\natexlab{}.
\newblock \showarticletitle{Remote Patient Activity Monitoring System by Integrating IoT Sensors and Artificial Intelligence Techniques}.
\newblock \bibinfo{journal}{\emph{Sensors}} \bibinfo{volume}{23}, \bibinfo{number}{13} (\bibinfo{year}{2023}), \bibinfo{pages}{5869}.
\newblock


\bibitem[Pang et~al\mbox{.}(2021)]%
        {pang2021technology}
\bibfield{author}{\bibinfo{person}{Carolyn Pang}, \bibinfo{person}{Zhiqin Collin~Wang}, \bibinfo{person}{Joanna McGrenere}, \bibinfo{person}{Rock Leung}, \bibinfo{person}{Jiamin Dai}, {and} \bibinfo{person}{Karyn Moffatt}.} \bibinfo{year}{2021}\natexlab{}.
\newblock \showarticletitle{Technology adoption and learning preferences for older adults: evolving perceptions, ongoing challenges, and emerging design opportunities}. In \bibinfo{booktitle}{\emph{Proceedings of the 2021 CHI conference on human factors in computing systems}}. \bibinfo{pages}{1--13}.
\newblock


\bibitem[Pasquetto et~al\mbox{.}(2020)]%
        {pasquetto2020tackling}
\bibfield{author}{\bibinfo{person}{Irene~V Pasquetto}, \bibinfo{person}{Briony Swire-Thompson}, \bibinfo{person}{Michelle~A Amazeen}, \bibinfo{person}{Fabr{\'\i}cio Benevenuto}, \bibinfo{person}{Nadia~M Brashier}, \bibinfo{person}{Robert~M Bond}, \bibinfo{person}{Lia~C Bozarth}, \bibinfo{person}{Ceren Budak}, \bibinfo{person}{Ullrich~KH Ecker}, \bibinfo{person}{Lisa~K Fazio}, {et~al\mbox{.}}} \bibinfo{year}{2020}\natexlab{}.
\newblock \showarticletitle{Tackling misinformation: What researchers could do with social media data}.
\newblock \bibinfo{journal}{\emph{The Harvard Kennedy School Misinformation Review}} (\bibinfo{year}{2020}).
\newblock


\bibitem[Pradhan et~al\mbox{.}(2020)]%
        {pradhan2020understanding}
\bibfield{author}{\bibinfo{person}{Alisha Pradhan}, \bibinfo{person}{Ben Jelen}, \bibinfo{person}{Katie~A Siek}, \bibinfo{person}{Joel Chan}, {and} \bibinfo{person}{Amanda Lazar}.} \bibinfo{year}{2020}\natexlab{}.
\newblock \showarticletitle{Understanding older adults' participation in design workshops}. In \bibinfo{booktitle}{\emph{Proceedings of the 2020 CHI Conference on Human Factors in Computing Systems}}. \bibinfo{pages}{1--15}.
\newblock


\bibitem[Rambhatla({[n.\,d.]})]%
        {rambhatlamaking}
\bibfield{author}{\bibinfo{person}{Sirisha Rambhatla}.} \bibinfo{year}{[n.\,d.]}\natexlab{}.
\newblock \showarticletitle{Making Canadian Healthcare Systems “AI READY”: What do we Need to Build AI-Powered Trustworthy Primary Healthcare Solutions?}
\newblock \bibinfo{journal}{\emph{Preface from Dr. Vivek Goel}} (\bibinfo{year}{[n.\,d.]}), \bibinfo{pages}{53}.
\newblock


\bibitem[Sabry et~al\mbox{.}(2022)]%
        {sabry2022machine}
\bibfield{author}{\bibinfo{person}{Farida Sabry}, \bibinfo{person}{Tamer Eltaras}, \bibinfo{person}{Wadha Labda}, \bibinfo{person}{Khawla Alzoubi}, \bibinfo{person}{Qutaibah Malluhi}, {et~al\mbox{.}}} \bibinfo{year}{2022}\natexlab{}.
\newblock \showarticletitle{Machine learning for healthcare wearable devices: the big picture}.
\newblock \bibinfo{journal}{\emph{Journal of Healthcare Engineering}}  \bibinfo{volume}{2022} (\bibinfo{year}{2022}).
\newblock


\bibitem[Sixsmith et~al\mbox{.}(2022)]%
        {sixsmith2022older}
\bibfield{author}{\bibinfo{person}{Andrew Sixsmith}, \bibinfo{person}{Becky~R Horst}, \bibinfo{person}{Dorina Simeonov}, {and} \bibinfo{person}{Alex Mihailidis}.} \bibinfo{year}{2022}\natexlab{}.
\newblock \showarticletitle{Older people’s use of digital technology during the COVID-19 pandemic}.
\newblock \bibinfo{journal}{\emph{Bulletin of Science, Technology \& Society}} \bibinfo{volume}{42}, \bibinfo{number}{1-2} (\bibinfo{year}{2022}), \bibinfo{pages}{19--24}.
\newblock


\bibitem[Vijaykumar(2020)]%
        {bmj}
\bibfield{author}{\bibinfo{person}{Santosh Vijaykumar}.} \bibinfo{year}{2020}\natexlab{}.
\newblock \bibinfo{booktitle}{\emph{Covid-19: Older adults and the risks of misinformation}}.
\newblock
\urldef\tempurl%
\url{https://blogs.bmj.com/bmj/2020/03/13/covid-19-older-adults-and-the-risks-of-misinformation/}
\showURL{%
Retrieved February 11, 2023 from \tempurl}


\bibitem[Vitak and Shilton(2020)]%
        {vitak2020trust}
\bibfield{author}{\bibinfo{person}{Jessica Vitak} {and} \bibinfo{person}{Katie Shilton}.} \bibinfo{year}{2020}\natexlab{}.
\newblock \showarticletitle{Trust, privacy, and security, and accessibility considerations when conducting mobile technologies research with older adults}. In \bibinfo{booktitle}{\emph{Mobile Technology for Adaptive Aging: Proceedings of a Workshop}}. National Academies Press, \bibinfo{pages}{1--20}.
\newblock


\bibitem[Wang et~al\mbox{.}(2021)]%
        {wang2021towards}
\bibfield{author}{\bibinfo{person}{Qiaosi Wang}, \bibinfo{person}{Koustuv Saha}, \bibinfo{person}{Eric Gregori}, \bibinfo{person}{David Joyner}, {and} \bibinfo{person}{Ashok Goel}.} \bibinfo{year}{2021}\natexlab{}.
\newblock \showarticletitle{Towards mutual theory of mind in human-ai interaction: How language reflects what students perceive about a virtual teaching assistant}. In \bibinfo{booktitle}{\emph{Proceedings of the 2021 CHI conference on human factors in computing systems}}. \bibinfo{pages}{1--14}.
\newblock


\bibitem[Wang and Hsu(2023)]%
        {wang2023integrating}
\bibfield{author}{\bibinfo{person}{Wei-Hsun Wang} {and} \bibinfo{person}{Wen-Shin Hsu}.} \bibinfo{year}{2023}\natexlab{}.
\newblock \showarticletitle{Integrating artificial intelligence and wearable IoT system in long-term care environments}.
\newblock \bibinfo{journal}{\emph{Sensors}} \bibinfo{volume}{23}, \bibinfo{number}{13} (\bibinfo{year}{2023}), \bibinfo{pages}{5913}.
\newblock


\bibitem[Wilson et~al\mbox{.}(2021)]%
        {wilson2021barriers}
\bibfield{author}{\bibinfo{person}{Jessica Wilson}, \bibinfo{person}{Milena Heinsch}, \bibinfo{person}{David Betts}, \bibinfo{person}{Debbie Booth}, {and} \bibinfo{person}{Frances Kay-Lambkin}.} \bibinfo{year}{2021}\natexlab{}.
\newblock \showarticletitle{Barriers and facilitators to the use of e-health by older adults: a scoping review}.
\newblock \bibinfo{journal}{\emph{BMC public health}}  \bibinfo{volume}{21} (\bibinfo{year}{2021}), \bibinfo{pages}{1--12}.
\newblock


\bibitem[Xie et~al\mbox{.}(2021)]%
        {xie2021going}
\bibfield{author}{\bibinfo{person}{Bo Xie}, \bibinfo{person}{Neil Charness}, \bibinfo{person}{Karen Fingerman}, \bibinfo{person}{Jeffrey Kaye}, \bibinfo{person}{Miyong~T Kim}, {and} \bibinfo{person}{Anjum Khurshid}.} \bibinfo{year}{2021}\natexlab{}.
\newblock \showarticletitle{When going digital becomes a necessity: Ensuring older adults’ needs for information, services, and social inclusion during COVID-19}.
\newblock In \bibinfo{booktitle}{\emph{Older Adults and COVID-19}}. \bibinfo{publisher}{Routledge}, \bibinfo{pages}{181--191}.
\newblock


\end{thebibliography}
\end{document}
\endinput